\documentclass[prd,twocolumn,superscriptaddress,nofootinbib,floatfix]{revtex4-2}
\usepackage{epsfig}
\usepackage{graphicx}
\usepackage[english]{babel}
\usepackage{hyphenat}
\usepackage{amsmath}
\usepackage{amssymb}
\usepackage{mathtools}
\usepackage{mathrsfs}
\usepackage{slashed}
\usepackage{epstopdf}
\usepackage{xcolor}
\usepackage{braket}
\usepackage{booktabs}
\definecolor{lcolor}{rgb}{0.,0.0,0.}
\definecolor{citcolor}{rgb}{0,0.,0.5}
\usepackage[breaklinks,colorlinks,urlcolor=blue,citecolor=blue,linkcolor=blue]{hyperref}
\usepackage{multirow}
\usepackage{ltablex}
\usepackage{soul}

\newcommand{\beq}{\begin{eqnarray}}
\newcommand{\eeq}{\end{eqnarray}}

\newcommand{\bem}{\begin{multline}}
\newcommand{\eem}{\end{multline}}
\newcommand{\beg}{\begin{gather}}
\newcommand{\eeg}{\end{gather}}

\newcommand{\nn}{\nonumber\\}

\newcommand{\ben}{\begin{eqnarray*}}
\newcommand{\een}{\end{eqnarray*}}

\newcommand{\secn}[1]{Section~1}
\newcommand{\appn}[1]{Appendix~1}

\long\def\comment#1{ }

\def\and{\quad\text{and}\quad}

\def\0{{\boldsymbol 0}}

\def\0{{\boldsymbol 0}}

\begin{document}

\title{Probing Celestial Energy and Charge Correlations \\ through Real-Time Quantum Simulations: Insights from the Schwinger Model}

\author{João Barata}
\email[]{jlourenco@bnl.gov}
\affiliation{Physics Department, Brookhaven National Laboratory, Upton, NY 11973, USA}

\author{Swagato Mukherjee}
\affiliation{Physics Department, Brookhaven National Laboratory, Upton, NY 11973, USA}

\begin{abstract}
Motivated by recent developments in the application of light-ray operators (LROs) in high energy physics, we propose a new strategy to study correlation functions of LROs through real-time quantum simulations. We argue that quantum simulators provide an ideal laboratory to explore the properties LROs in lower-dimensional quantum field theories (QFTs). This is exemplified in the 1+1-d Schwinger model, employing tensor network methods, focusing on the calculation of energy and charge correlators. Despite some challenges in extracting the necessary correlation functions from the lattice the methodology used can be extended to real quantum devices. 
\end{abstract}

\maketitle

In recent years there has been a renewed effort towards describing high energy scattering processes in terms of the intrinsic correlations present in the final state at asymptotic spatial distances. In QFTs, these properties can be properly studied in terms of time-integrated correlation functions of conserved current operators living on a celestial sphere -- these correspond to the theoretical realizations of detectors used in real high energy experiments. This class of correlation functions has been studied to great depth in conformal field theories (CFTs)~\cite{Kravchuk:2018htv,Kologlu:2019mfz,Chang:2020qpj,Korchemsky:2015ssa,Belitsky:2014zha,Belitsky:2013xxa,Zhiboedov:2013opa,Cordova:2017zej,Cordova:2017dhq,Manenti:2019kbl,Korchemsky:2021htm}, providing a wealth of techniques and physics insights. These developments have been translated to quantum chromodynamics (QCD), establishing a remarkable connection between theory~\cite{Lee:2022ige,Chen:2021gdk,Chen:2020vvp,Dixon:2019uzg,Basham:1977iq,Tkachov:1995kk,Sveshnikov:1995vi,Korchemsky:1999kt}, phenomenology~\cite{Chen:2024nfl,Holguin:2024tkz,Komiske:2022enw,Moult:2018jzp,Cao:2023oef,Riembau:2024tom,Bossi:2024qho,Andres:2024ksi,Yang:2023dwc,Barata:2023bhh,Li:2021txc,Barata:2023zqg,Singh:2024vwb}, and experiment~\cite{CMS:2024ovv,CMS:2024mlf}.\footnote{See also~\cite{Chicherin:2024ifn,Belitsky:2013bja,Belitsky:2013ofa,Chicherin:2024ifn} for related developments in $\mathcal{N}=4$ SYM.}

In contexts closer to QCD, the most common realization of this program is in terms of the $N$-point energy correlators (E$N$Cs), which measure the correlations between asymptotic energy flows~\cite{Hofman:2008ar,Basham:1977iq,Basham:1979gh,Basham:1978bw}. E$N$Cs can be defined in terms of the light-transformed energy momentum tensor (EMT), $T^{\mu \nu}$:
\begin{align}\label{eq:def_Ec}
 \mathcal{E}(\vec n) = \lim_{r\to \infty} r^{d-2} \int_0^\infty dt\, n^i T^{0i}(t,r\, \vec n) \, ,
\end{align}
where $\mathcal{E}$ denotes the LRO corresponding to the local EMT operator, depending only on the unit vector $\vec n$ labeling a point on the celestial $d-1$ dimensional sphere. Using Eq.~\eqref{eq:def_Ec}, E$N$Cs are then correlation functions of the generic form
\begin{align}\label{eq:ENC_general}
 \frac{\langle \Omega| K^\dagger \mathcal{E}(\vec n_1) \mathcal{E}(\vec n_2) \cdots \mathcal{E}(\vec n_{N})  K|\Omega\rangle}{ \langle\Omega|  K^\dagger K |\Omega\rangle }  \, ,
\end{align}
where $|\Omega\rangle$ denotes an equilibrium state of the theory, e.g. the vacuum, and the operator $K$ generates a local excitation. In this work we also will consider charge correlations, obtained by replacing $T^{0i}$ in Eq.~\eqref{eq:def_Ec} with an appropriate conserved current operator $J^i$.

Correlation functions of the form of Eq.~\eqref{eq:ENC_general} have been studied, in QCD, to high perturbative precision in certain kinematical limits for $N\leq 3$~\cite{Moult:2018jzp, Chen:2019bpb,Yang:2022tgm,Dixon:2019uzg}. Accessing the properties of E$N$Cs in full kinematics, for large values of $N$ and beyond the perturbative region is a challenging task,\footnote{See~\cite{Chen:2024nyc,Lee:2024esz,Schindler:2023cww,Dokshitzer:1999sh,Lee:2006nr,Komiske:2022enw} for studies on non-perturbative effects on E$N$Cs.} but one which could provide further important information about the theory. Although in QCD many of these aspects can be addressed by the use of dedicated Monte-Carlo simulations, such approaches lack a direct connection to the underlying field theory  problem; furthermore attempts to study E$N$Cs from euclidean lattice field theory simulations are not possible due to the Lorentzian nature of the light-transform in Eq.~\eqref{eq:def_Ec}. Thus, new methods to study LRO's correlation functions from first-principles would considerably enlarge the power of the current theoretical toolkit set.

\begin{figure*}
    \centering
    \includegraphics[width=.95\textwidth]{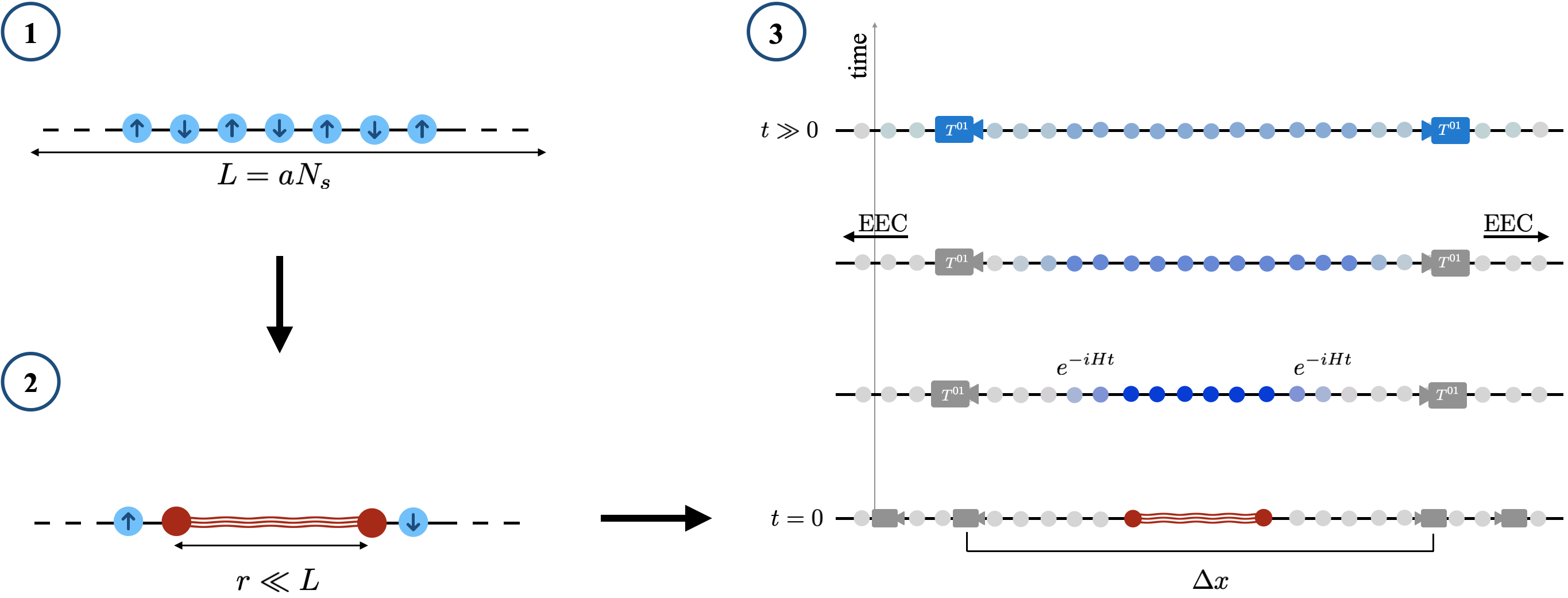}
    \caption[width=.95\textwidth]{Protocol used to extract the correlation functions studied in the main text from the lattice: \textbf{1)} we first prepare the ground state of the theory, here illustrated in the strong coupling limit; \textbf{2)} the system is then quenched by inserting a local expanding electric field, with maximum size $r \ll L$; \textbf{3)} once the external electric field reaches its maximum length, it is turned off; detector \textit{cameras} displaced by $\Delta x$ are then placed at the edge of the lattice and the system evolves naturally under $H$. In \textbf{3)}, we sketch how the initial energy deposited in the center of the lattice (dark blue) diffuses over time (lighter blue) away from the center. Once the energy flow reaches a set of detectors at $t\gg0$, they register a signal (blue cameras), as illustrated in Fig.~\ref{fig:EEC}. The EEC limit would be realized by taking the detectors to large separations and integrating over time, while taking a continuum limit.  
    }
    \label{fig:scheme}
\end{figure*}

In this work, we argue that real time quantum simulations can provide such a new direction, both in the context of gauge theories and for CFTs, serving as a laboratory to test theoretical predictions. While this program could be realized in the near-to-mid future, taking into account the development of quantum technologies and devices for the study of QFTs~\cite{Bauer:2022hpo,DiMeglio:2023nsa,Preskill:2018jim,Halimeh:2023lid,Shaw:2020udc,Hardy:2024ric}, here we will illustrate our approach in the Schwinger model~\cite{Schwinger:1962tp}, QED in $1+1$-d, making use of tensor network methods~\cite{Banuls:2019bmf}. These are ideally suited for quantum simulations of lower dimensional systems and can be naturally mapped quantum computer based approaches. In the $A^0=0$ gauge, this theory is defined by the Hamiltonian:
\begin{align}\label{eq:H_Schwinger}
	H&=  \int dx \, \frac{E^2(x)}{2} \nn 
 &+ \bar  \psi(x) (-i \gamma^1 \partial_1 +g \gamma^1 A_1(x) +m ) \psi(x)\, ,
\end{align}
with $\psi$ a two component, single flavor massive fermionic field. Due to dimensionality, the properties of correlations of LROs in this model are quite distinct from what is seen in $3+1$-d. Firstly, in $1+1$-d the celestial sphere consists of a set of two disconnected points; as a result there is, for example, no notion of a light-ray operator product expansion (OPE), which plays a big role in the study of E$N$Cs, even though there is a local OPE. Further, since there are only two spatial asymptotic positions where one can place detectors, it only makes sense to consider single and two body correlations functions of the form 
\begin{align}\label{eq:def_ENC}
&C_{1,k}^{\mu\nu}(q)  =  \int dt \, dx \, e^{iq^\alpha x_\alpha} \left\langle J^\mu(t,x)  \mathcal{E}_k(\{R,L\})  J^\nu(0) \right \rangle \, , \nn 
&C_{2,k}^{\mu\nu}(q)  =  \int dt \, dx \, e^{iq^\alpha x_\alpha} \left\langle J^\mu(t,x)  \mathcal{E}_k(L) \mathcal{E}_k(R)  J^\nu(0) \right \rangle  \, .
\end{align}
Here the electromagnetic currents $J^\mu = \bar \psi \gamma^\mu \psi$ act on the vacuum state, generating a local injection of momentum $q$, which puts the system out of equilibrium, following Eq.~\eqref{eq:def_ENC}, while the $R$, $L$ denotes the insertion of the detector at asymptotic positive ($R$) or negative ($L$) spatial locations.\footnote{We note that to obtain the projection that would contribute to physical processes, one should contract $C_{1/2}$ with the appropriate polarization vectors; however in the Schwinger model the gauge field is strictly non-propagating, and thus such a contraction is vanishing. Secondly, the correlators should be normalized by the expectation value of the current insertions without measurement, as in Eq.~\eqref{eq:def_ENC}.} The index $k=1,2,3, \cdots$ in the LRO takes into account the possibility to measure integer powers of the operator~\cite{Caron-Huot:2022eqs}. Translation invariance and energy conservation determine $C_{1,k}$, since after injecting momentum $q$ into the system, equal amounts of momentum have to flow through both $R$ and $L$. The same statement holds for charge correlators, provided one works in a fixed charge sector set by Gauss's law. Although these constraints also affect $C_{2,k}$, they should not fully determine it, especially if one takes $k>1$.

To study the EEC in the Schwinger model one could use the fact that the theory, in the charge zero sector, is dual to a quantum Sine-Gordon model~\cite{PhysRevD.11.3594} via a bosonization map. In particular, in the limit of massless fermions, the dual theory is that of free massive bosons, with both being connected by the identity~\cite{PhysRevD.11.2088,PhysRevD.11.3594}
 \begin{align}
  \bar \psi \gamma^\mu \psi = \frac{1}{4\pi} \varepsilon^{\mu \nu} \partial_\nu \phi \, , 
\end{align}
where $\phi$ is a bosonic field with mass $m_\phi=g/\sqrt{\pi}$. 
Combining this with Eqs.~\eqref{eq:def_ENC} and that the off-diagonal component for the bosonic EMT reads $T^{0i}= \partial_0 \phi \partial_i \phi$, one can obtain for example:
\begin{align}\label{eq:EEC_masless}
 C_{1,1}^{\mu \nu} &=  \frac{\lim_{y\to \infty}}{16\pi^2}  \int_0^\infty ds\, \varepsilon^{\mu \alpha}  \varepsilon^{\nu \beta} \nn 
 &\times\partial^\alpha_x \partial^\beta_{x_0}  \partial^0_{y_1} \partial^1_{y_2}\langle \phi(x) \phi(y_1) \phi(y_2) \phi(x_0) \rangle_{y_1=y_2=y} \, .
\end{align}
Thus, computing the E$N$C boils down to the computation of $2(N+1)$-body Wightman functions of bosonic fields; in the free limit the evaluation of such expectation values is immediate through Wick's theorem (for the creation/annihilation operators). However for massive fermions, the dual theory is interacting ($g_\phi \propto m$), and the computation of the above correlation is non-trivial, see e.g.~\cite{Kukuljan:2018whw,Nomura:1995tq}. As a result, even in this simple model, the analytical computation of E$N$Cs beyond perturbation theory is far from obvious.

Alternatively, one can study this model on the lattice, where the above mentioned real-time simulation methods can be directly applied. To that end, we discretize space into $N_s/2$ sites, with a physical lattice spacing $2a$. Following the Kogut-Susskind prescription~\cite{Kogut:1974ag,Susskind:1976jm} for fermions,\footnote{See~\cite{Barata:2024bzk} for further details following the same conventions.} the continuum two component fermionic field operator is mapped to the staggered single component field $\chi_n$:
\begin{align}\label{eq:help_1}
   &\psi(x)  \to \frac{1}{\sqrt{a}}\begin{pmatrix}
		\chi_{2 \tilde n}  \\
		\chi_{2 \tilde  n-1}
	\end{pmatrix}\, , \nn 
  &\partial_1 \psi(x) \to  \frac{1}{a\sqrt{a}}\begin{pmatrix}
		\chi_{2 (\tilde n+1)} -\chi_{2 \tilde n}\\
		\chi_{2 \tilde  n-1} - 	\chi_{2 (\tilde  n-1)-1} 
  \end{pmatrix}\, .
\end{align}
Here $1<\tilde n<N_s/2$ is a physical lattice index, while we use $1<n<N_s$ for the computational lattice. Combining Eq.~\eqref{eq:help_1} with a Jordan-Wigner transform (JWt), $\chi_n =   S_n \sigma^-_n$, where the string operator is defined as $ S_n\equiv   \prod_{ k< n}[(-i)\sigma^z_k ]$, the Hamiltonian in Eq.~\eqref{eq:H_Schwinger} can be explicitly mapped to the spin-chain
\begin{align}\label{eq:Spin_Hamiltonian}
	H &= 	\frac{g^2a}{2} \sum_{n=1}^{N_s-1} \left[\frac{1}{2}\sum_{k=1}^n (\sigma^z_k+(-1)^k) \right]^2  +\sum_{n=1}^{N_s}   (-1)^n m\frac{\sigma^z_n}{2} \nn
	&+\frac{1}{2a} \sum_{n=1}^{N_s-1}  \sigma^+_n \sigma^-_{n+1} + \sigma^+_{n+1}  \sigma^-_n  \, .
\end{align}
Here $\sigma^{\{x,y,z,+,-\}}_n$ denotes the standard Pauli matrices acting on the computational site $n$. The first term in Eq.~\eqref{eq:Spin_Hamiltonian} is obtained after explicitly integrating out the gauge degrees of freedom, using Gauss's law, such that an all-to-all potential emerges. The energy and charge LROs can also be constructed by latticizing Eq.~\eqref{eq:def_ENC}; for the energy case we use that the off-diagonal part of the EMT in Schwinger model can be written as
\begin{align}
 T^{01}(x) =    \frac{i}{2} \left( \bar \psi (x) \gamma^0 D^1 \psi(x) - D^1 \bar \psi(x) \gamma^0 \psi(x) \right)\, ,
\end{align}
where $D^\mu$ denotes the covariant derivative, and we have chosen to use a non-symmetric form, instead of the symmetrized Belifante EMT. Both operators only differ by a total derivative, and thus they yield the same expectation values. Nonetheless, we note the practicality of the present form which only involves spatial derivatives, while the Belifante representation includes time derivatives which require the operator $[H,\psi]$.\footnote{We note that the construction of the EMT operator on the lattice in higher dimensions is more evolved and the current arguments are limited to $1+1$-d, see~\cite{Cohen:2021imf} for a detailed discussion.} Finally, the latticized EMT can be obtained by making use of the correspondence
\begin{align}
\bar \psi(t,z) \gamma^0 D^1 \psi(t,z) &\to  \frac{1}{a^2} e^{iHt}  \Bigg[\chi_{2 \tilde n}^\dagger \chi_{2\tilde n+2}- \chi^\dagger_{2\tilde n}  \chi_{2\tilde n} \nn 
 &+ \chi^\dagger_{2\tilde n-1} \chi_{2 \tilde n-1} - \chi^\dagger_{2\tilde n-1} \chi_{2 \tilde n-3}\Bigg] e^{-iHt}\, ,
\end{align}
where we have performed a residual gauge transformation $\psi(x) \to U(y<x)\psi(x)$, where $U$ denotes the link operators to the left of the point $x$, to explicitly remove the dependence on the gauge field, as was done at the level of the Hamiltonian to obtain Eq.~\eqref{eq:Spin_Hamiltonian}. A similar treatment can be employed for the charge operator, for which one finds the relation, after a JWt, 
\begin{align}
   J^1(x,t) \to e^{iHt}\frac{i}{a}\left(\sigma^+_{2\tilde n+1}\sigma^-_{2\tilde n} -\sigma^+_{2\tilde n}\sigma^-_{2 \tilde n+1} \right) e^{-iHt}\, .
\end{align}

Using the latticized Schwinger model and the appropriate representations of the detector operators, we can then study the correlation functions of the form of Eq.~\eqref{eq:ENC_general} numerically. To that end, we implement the protocol illustrated in Fig.~\ref{fig:scheme}. First, we prepare the ground state of the Schwinger model at finite $m$ and $g$; this generates the state $|\Omega\rangle $ in Eq.~\eqref{eq:ENC_general}. Second, we insert an expanding electric field\footnote{The form of the quench is not particularly important, as long as it leads to an increase in energy density and it is localized in space. It is simple to check that this is indeed achieved by this quench~\cite{PhysRevD.10.732,Hebenstreit:2013baa}, see also~\cite{Florio:2024aix,Florio:2023dke,Barata:2023jgd,Banerjee:2012pg,Farrell:2024mgu} for related studies.} in the center of the lattice and let the system evolve for a short time, such that the total string length $r$ is much smaller than the total system length $L=N_s a$. This external electric field, when acting on the vacuum, puts the system out of equilibrium, playing the same role as the external electromagnetic currents entering Eqs.~\eqref{eq:ENC_general}. However, applying this external field is computationally less complex than computing the expectation values of LROs with external currents, while still achieving the same qualitative goal of injecting energy into to the system. Finally, after having prepared this locally out of equilibrium state, we insert local operators on both sides of the lattice with a $\Delta x\sim \mathcal{O}(L)$ spatial separation, and let the system naturally evolve under $H$. In the case where the inserted operator is the EMT, we can then directly extract:
\begin{align}
    C^{\rm lat.}_{2,k}(\Delta x,t,t') &\equiv \Big\langle e^{iHt}  [T^{01}(-\Delta x/2)]^k e^{iH(t'-t)} \nn 
   &\times [T^{01}(\Delta x/2)]^k  e^{-iHt'}\Big\rangle_{\rm quench} \, ,
\end{align}
where $t,t'$ denote the time variable with respect to the left and right detectors. To obtain the EEC, one needs to integrate over time, place the detectors at increasingly large $\Delta x$, and take the continuum limit, i.e. $\mathcal{C}_{2,k}=\lim_{\Delta x \to \infty}\int_{t,t'}C^{\rm lat.}_{2,k}(\Delta x,t,t')$.

\begin{figure}
    \centering
    \includegraphics[width=.45\textwidth]{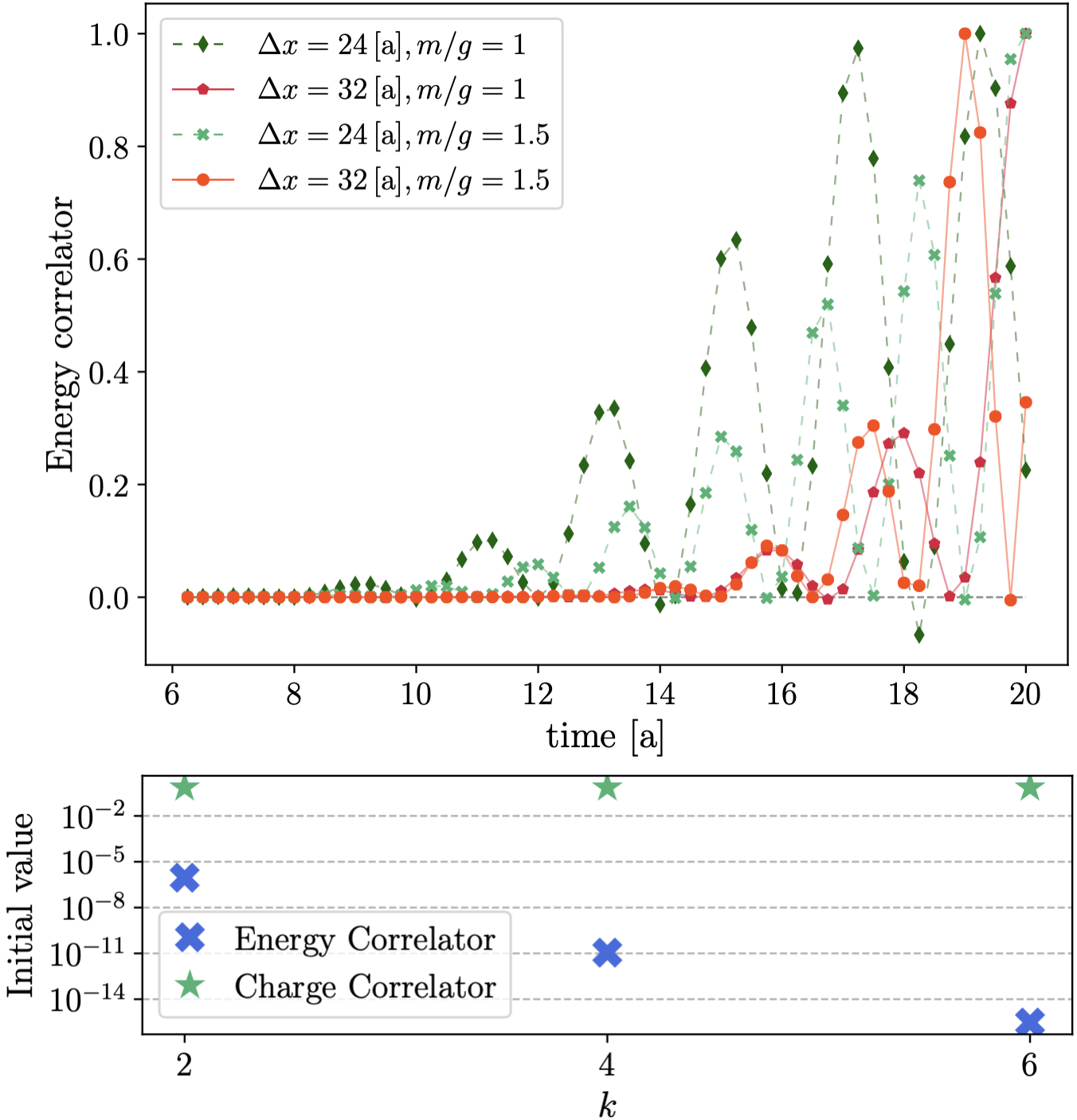}
    \caption[width=.95\textwidth]{\textbf{Top:} Evaluation of $C_{2,1}^{\rm lat.}$ as a function of time, for two values of $m/g=1,\, 1.5$ and a shorter ($\Delta x= 24\, a$) and larger ($\Delta x= 32\, a$) spatial separations of the detectors. To collect all curves, we have normalized the $y-$axis to the maximum extracted value each data set, in this time domain. We numerically checked that after such a renormalization of the curves the results for the charge correlator show an identical behavior. \textbf{Bottom:} Extraction of the initial value, i.e. at times before the detector detects the signal, for the energy and charge correlators with power $k$, using $m/g=1$. For odd values of $k$ we find that this quantity exactly vanishes, as illustrates above for the $k=1$ case. For even values of $k$, the charge correlator gets a non-vanishing positive value which does not depend on $k$. This behavior differs from the energy correlator, where the initial value decreases with $k$, as illustrated. Note that here we have normalized the expectation values of the energy correlator to $\langle H^{2k} \rangle_{\rm quench}$.}
    \label{fig:EEC}
\end{figure}

In Fig.~\ref{fig:EEC} we show the evaluation of $C^{\rm lat.}_{2,k}(\Delta x,t,t)$.\footnote{Note that here we take $t=t'$ and we do not perform the time integrals. The reasoning for this is twofold. First, the EEC in two dimensions is a number, and thus having the time differential correlator provides more information. Secondly, taking different times for the two operator insertions makes the simulation technically more costly; since we do not compute the full EEC looking the equal time correlator seems sufficient.} The numerical simulations are performed using a MPS tensor network, implemented using the \texttt{Itensor} package~\cite{Fishman:2020gel}. We use a lattice with $N_s=60$ lattice sites, while state preparation of the initial state in Fig.~\ref{fig:scheme} is done using the native density matrix renormalization group (DMRG)~\cite{White:1992zz,White:1993zza} algorithm, and the real-time evolution is implemented via the time-dependent variational principle (TDVP)~\cite{Haegeman:2011zz, Haegeman:2016gfj} algorithm implemented in \texttt{Itensor}. The maximal spatial size of the external electric is taken to be $r = 11 \, a $. This generates a variation of the total energy of the system, i.e. $(\langle H \rangle_{\rm vacuum}-\langle H \rangle_{\rm quench})/\langle H \rangle_{\rm vacuum} \approx \mathcal{O}(0.1 -1  \, \%)$, compared to the vacuum state and after normal ordering of the Hamiltonians.

In Fig.~\ref{fig:EEC} (top) we show the late time behavior of $C^{\rm lat.}_{2,k}(\Delta x,t,t)$, for two values of spatial separation of the detectors and two values of $m/g$. At times $t<6 \, a$, the correlator exactly vanishes. This is in agreement with the naive expectation from the right hand side illustration in Fig.~\ref{fig:scheme}, where a finite amount of time is required for the \textit{signal} to propagate from the initial localized perturbation to the detectors. Comparing the starting time for the variations of the energy correlator for the two values of $\Delta x$, one concludes that the momentum spread happens close to the (lattice) speed of light, i.e. $\Delta(\Delta x)/\delta t \sim \mathcal{O}(0.5-1)$ where we roughly estimate the time interval by comparing the first peaks of the dark green and red data sets. The oscillatory-like behavior seen after the first peaks is sensitive to the lattice edge (due to the small gap with respect to the \textit{spatial infinity} where the detectors are placed), and thus the extracted results there are not physical. To overcome these issues, one should not only ensure that $\Delta x \sim N_s a$ but also that $ |x_{\rm detector}| < N_s a/2$, with $x_{\rm detector}$ the spatial position of a detector; nonetheless the realization of such a limit is numerically hard to achieve due to the need to time evolve the full quantum state on a larger lattice.

A similar exercise to that shown on the top panel of Fig.~\ref{fig:EEC} was also carried out for the charge correlator. Remarkably, we found for several parameters sets, that if one normalizes the curves as in Fig.~\ref{fig:EEC} -- i.e. by dividing all points in each data set by the maximum value in the considered time domain -- that the energy and charge correlators have a very close behavior. This qualitative observation agrees with the following simple picture: \textbf{1)} the evolution of the correlators is constrained by conservation of energy and conservation of total charge;\footnote{We work in a charge zero sector with no net momentum.} \textbf{2)} as energy flows outwards, left and right, it must do so in a balanced way since there are no other directions; \textbf{3)} a similar statement must be true about charge, since any imbalance would violate Gauss's law; \textbf{4)} this would support the \textit{naive} picture that an electron flying to the left of the lattice must be compensated by an equivalently charged \textit{state} flying to the right of the lattice. Of course, the explanation is more complicated since the axial and vector currents are related non-trivially in $1+1$-d, where $J^\mu\propto \varepsilon^{\mu \nu}  J_5^\nu$, and thus the true degrees of freedom of the system are not fermionic. These aspects require a more detailed understanding of the particular structure of the underlying quantum state. Conversely, they can also be studied by computing the correlations functions of the axial $J_5^1$ current, and mixed correlators, e.g. $\langle T^{01} J^1 \rangle_{\rm quench}$, which we leave for future work.

On the bottom of Fig.~\ref{fig:EEC}, we show the results for the \textit{initial value} of the energy and charge correlator, i.e. the correlator value for times earlier than the observation of the first peak, for several values of $k$. The odd values are not shown since for the those cases the extracted value is exactly vanishing. More, we observed that for the odd values of $k$ the behavior in time of the energy and charge correlation functions is very close to what is seen for $k=1$. These statements do not hold for even $k$, where we observe that the charge correlator acquires a finite initial value, which does not depend on $k$, while the energy correlator acquires a positive value which grows with $k$.\footnote{Note that in the figure the evolution with $k$ has the opposite trend. This is because we normalize each point by $\langle H^{2k} \rangle_{\rm quench}$, which grows considerably faster than the evolution of the \textit{bare} correlator.} This indicates that in the massive Schwinger model these two currents can not be identified (as one should expect), complementing the above observations. Further, one would expect that these deviations are sensitive to the fluctuations of the energy and charge asymptotic transport, which can not be fully constrained by global energy and charge conservation.

In this manuscript we have provided a first discussion on the study of LROs' correlations from real-time lattice simulations. This new approach can be potentially used to explore non-perturbative features of E$N$Cs in gauge theories and complements existing perturbative methods. Of course, the realization of such a program is tightly dependent on the development of large scale quantum computers, which are not available at the moment. This is particularly important for the extraction of LROs' correlations functions since the measurements must be performed at large spatial separations. Combined with the need to take continuum limits and the large number of degrees of freedom necessary to implement gauge theories in quantum devices, such an endeavor is extremely resource intensive. Hopefully continuum and large volume extrapolations may be possible in mid to long term future, using more advanced quantum devices. Perhaps more interesting for near term implementations is the study of $2+1$-d field theories where the behavior near critical points of LROs' can be theoretically described; a natural example is the three dimensional critical Ising model, which can be more easily realized in quantum devices and (large) tensor networks, when compared to gauge theories. We leave such a study for future work, see~\cite{ziegler1993energyenergy,Giuliani_2012} for related discussions. From the point of view of quantum simulation of QFTs, the study of LROs is of critical importance since they provide a proper way to introduce the notion of \textit{detector}, a necessary element in the study of high energy physics using  lower dimensional models, see e.g.~\cite{Bauer:2022hpo}.

\textit{Acknowledgments}: We thank I. Moult, K. Lee, M. Riembau, A. V. Sadofyev, R. Szafron, and H. Zhu for useful discussions. We thank Henry Lamm for useful comments on the form of the energy momentum tensor on the lattice. This material is based upon work supported by the U.S. Department of Energy, Office of Science, Office of Nuclear Physics and National Quantum Information Science Research Centers, Co-design Center for Quantum Advantage (C2QA) under contract number DE-SC0012704. The work of S.M. is partially supported by Laboratory Directed Research and Development (LDRD) funds from Brookhaven Science Associates.

\bibliography{references.bib}

\end{document}